\documentclass[12pt,a4paper]{article}
\usepackage{amsmath,amsthm,amsfonts,amssymb,cite}
\usepackage[dvips]{graphicx}

\DeclareSymbolFont{bletters}{OML}{cmm}{bx}{it}
\DeclareMathSymbol{\bla}{\mathord}{bletters}{'025}
\DeclareMathSymbol{\bmu}{\mathord}{bletters}{'026}
\DeclareMathSymbol{\bth}{\mathord}{bletters}{'022}
\DeclareMathSymbol{\bfI}{\mathord}{bletters}{"49}
\DeclareMathSymbol{\bdl}{\mathord}{bletters}{"0E}
\DeclareMathSymbol{\bDl}{\mathord}{bletters}{"001}
\def \bpi{\boldsymbol\pi}
\def \bphi{\boldsymbol\phi}

\def \si{\sigma}
\def \la{\lambda}
\def \be{\beta}
\def \ta{\theta}
\def \dl{\delta}

\def \Dl{\Delta}
\def \CA{\mathcal A}
\def \CM{\mathcal M}
\def \CN{\mathcal N}
\def \CV{\mathcal V}
\def \CU{\mathcal U}

\def \CP{\mathcal P}
\def \CZ{\mathcal Z}
\def \CT{\mathcal T}
\def \BC{\mathbb{C}}
\def \BR{\mathbb{R}}
\def \BZ{\mathbb{Z}}
\def \BI{\mathbb{I}}
\def \inf{{\rm{SA}}}

\begin{document}
\title{
$${}$$\\
{\bf The Correlation Functions} \\
{\bf of the ${\bf XXZ}$ Heisenberg Chain for}\\
{\bf Zero or Infinite Anisotropy and}\\
{\bf Random Walks of Vicious Walkers{\footnote{Extended talk at the Conference ``Conformal Field Theory, Integrable Systems, and Liouville Gravity'' (Chernogolovka, June 30 -- July 2, 2009)}}
{\footnote{Partially supported by RFBR (No.~07-01-00358) and by the Russian Academy of Sciences program `Mathematical Methods in Non-Linear Dynamics'}} } }
\author{
$${}$$\\
{\bf N.~M.~Bogoliubov$^\star$, C.~Malyshev$^\diamond$}\\[0.5cm]
{\small Steklov Mathematical Institute,
St.-Petersburg Department, RAS}\\
{\small Fontanka 27, St.-Petersburg, 191023, Russia} \\
[0.0cm]
$^\star$ e-mail: {\it bogoliub@pdmi.ras.ru}\\
\,\,$^\diamond$ e-mail: {\it malyshev@pdmi.ras.ru} }

\date{}

\maketitle

\begin{abstract} \noindent
The $XXZ$ Heisenberg chain is considered for two specific limits of the anisotropy parameter: $\Dl\to 0$ and $\Dl\to -\infty$. The corresponding wave functions are expressed by means of the symmetric Schur functions. Certain expectation values and thermal correlation functions of the ferromagnetic string operators are calculated over the base of $N$-particle Bethe states. The thermal correlator of the ferromagnetic string is expressed
through the generating function of the lattice paths of random walks of vicious walkers. A relationship between the expectation values obtained and the generating functions of strict plane partitions in a box is discussed. Asymptotic estimate of the thermal correlator of the ferromagnetic string is obtained in the limit of zero temperature. It is shown that its amplitude is related to the number of plane partitions.
\end{abstract}

\leftline{\emph{{\bf Keywords:}} XXZ Heisenberg chain, Schur functions, random walks, plane partitions}

\thispagestyle{empty}

\newpage

\section{Introduction}

\subsection{XXZ Heisenberg chain}

A system of spin \textit{1/2} particles occupying sites of one-dimensional lattice, widely known as the quantum \textit{XYZ Heisenberg chain} \cite{hg}, has attracted considerable attention both in theoretical and mathematical physics, and it has been thoroughly investigated for a long time \cite{bt, yy1, yy2, yy3, lib, bax, god}. The \textit{Quantum Inverse Scattering Method}, developed for solving the integrable models of quantum field theory and statistical physics \cite{f1, kuls}, has been also used to investigate the $XYZ$ Heisenberg chain \cite{ft, ft2}. An important special case of the $XYZ$ model, so-called, $XXZ$ spin chain also attracts considerable attention \cite{vk, vk1, ess, ml1, ml2}. The Hamiltonian of the $XXZ$ magnet has been diagonalized by the coordinate \textit{Bethe Ansatz Method} in \cite{yy1, yy2, yy3}. The \textit{Algebraic Bethe Ansatz} has been used in \cite{ft} to solve the $XXZ$ model. The problem of calculation of the correlation functions of the model in question in the framework of the Algebraic Bethe Ansatz has required serious efforts \cite{vk, vk1, ks, KBI1, KBI2}.

The random walks is a classical problem both for combinatorics and statistical physics. The problem of enumeration of the paths made by the, so-called, \textit{vicious} walkers on the one-dimensional
lattice has been formulated and investigated in details by Fisher~\cite{1}. The problem mentioned still continues to attract considerable attention both of physicists and mathematicians~\cite{2, 3, 4, 5, 6, 7, 8, 9, 10, 11, 12, 13, 14}.

Random walks on one-dimensional periodic lattice can be related to the correlation functions of the $XX$ Heisenberg magnet. Certain operator averages taken over the ferromagnetic state of the $XX$ model play a role of the generating functions of the number of paths traced by the vicious walkers. The problem of enumeration of trajectories of vicious walkers by means of the correlation functions of the $XX$ model has been studied in the series of papers \cite{b1, b2, b3, b4}. The approach of Refs. \cite{b1, b2, b3, b4} will be further explored in the present paper. We shall consider two limits of the $XXZ$ spin chain from the point of view of random walks of vicious walkers, as well as from a viewpoint of enumeration of boxed plane partitions.

Let us begin with the $XXZ$ model defined on one-dimensional lattice consisting of $M+1$ sites labeled by elements of the set $\CM\equiv\{0 \le k \le M,\,k \in\BZ\}$, $\,M+1=0\pmod{2}$. The corresponding spin Hamiltonian is defined, in absence of external magnetic field, as follows:
\begin{equation}
{\hat H}_{XXZ}=-\frac 12\sum_{k=0}^M (\si_{k+1}^{-}\si_k^{+} + \si_{k+1}^{+}\si_k^{-}+\frac\Dl 2\,(\si_{k+1}^z\si_k^z-1))\,, \label{xxzham}
\end{equation}
where the parameter $\Dl\in\BR$ describes the internal anisotropy of the model. For instance, the choice $\Dl=\pm 1$ corresponds to, so-called, isotropic $XXX$ spin chain solved in \cite{bt}. The local spin operators $\si^\pm_k = \frac12(\si^x_k\pm i\si^y_k)$ and $\si^z_k$, dependent on the lattice argument $k\in\CM$, are defined as $(M+1)$-fold tensor products as follows:
\begin{equation}
\si^{\#}_k=\si^0\otimes \dots\otimes \si^0\otimes
\underbrace{\si^{\#}}_{k} \otimes\,\si^0 \otimes \dots\otimes
\si^0\,,
\label{xxzham1}
\end{equation}
where $\si^0$ is $2\times 2$ unit matrix, and $\si^{\#}$ at k$^{th}$ place denotes a Pauli matrix, $\si^{\#} \in {\mathfrak{su}}(2)$ (superscript $\#$ implies either $x, y, z$ or $\pm$). Therefore, the spin operators act over the state-space ${\mathfrak H}_{M+1}$ given by the tensor product of $M+1$ copies of the linear spaces ${\mathfrak h}_{k} \equiv {\BC}^2$: ${\mathfrak H}_{M+1} = \bigotimes \limits_{k=0}^{M} {\mathfrak h}_{k}$. The commutation rules for the spin operators are given by the relations:
\[
[\,\si^+_k,\si^-_l\,]\,=\,\dl_{k,l}\,\si^z_l\,,\quad
[\,\si^z_k,\si^\pm_l\,]\,=\,\pm 2\,\dl_{k,l}\,\si^{\pm}_l\,.
\]
The linear space ${\BC}^2$ is spanned over the spin ``up'' and ``down'' states ($\mid\uparrow\rangle$ and $\mid\downarrow \rangle$, respectively) providing a natural basis so that
\[
\mid\uparrow\rangle\equiv \Bigl(
\begin{array}{c}
1 \\
0
\end{array}
\Bigr)\,,\qquad\mid\downarrow\rangle\equiv \Bigl(
\begin{array}{c}
0 \\
1
\end{array}
\Bigr)\,.
\]
The space ${\mathfrak H}_{M+1}$ is spanned over of the state-vectors $\bigotimes\limits_{k=0}^M \mid\!\! s\rangle_{k}$ , where $s$ implies either $\uparrow$ or $\downarrow$. The periodic boundary conditions $\si^{\#}_{k+(M+1)}=\si^{\#}_k$ are imposed.

To represent $N$-particle state-vectors of the model, $\mid\!\Psi_N(u_1,\dots,u_N)\rangle$, let the sites with spin ``down'' states be labeled by the coordinates $\mu_i$, $1\leq i\leq N$. These coordinates form a strict partition $\boldsymbol{\mu}\equiv (\mu_1, \mu_2,\,\dots\,, \mu_N)$, where $M\geq\mu_1>\mu_2>\,\dots\,> \mu_N \geq 0$. There is a correspondence between each partition and an appropriate sequence of zeros and unities of the form: $\bigl\{e_k\equiv e_k(\boldsymbol{\mu})\bigr\}_{k\in \CM}$, where $e_k \equiv \dl_{k, \mu_n}$, $1 \leq n\leq N$. This correspondence enables another convenient expression for the strict partitions: $\bmu = \left(M^{e_M}, \ldots , 1^{e_1}, 0^{e_0} \right)$. It is meant here that any non-negative integer $k\in\CM$ appears $e_k$ times in the present configuration, and the condition $\sum\limits_{k=0}^M e_k = N$ is respected. The Hamiltonian (\ref{xxzham}) is diagonalized \textit{via} the ansatz:
\begin{equation}
\mid\!\Psi_N({\textbf u})\rangle \,=\,\!\!
\sum_{\{e_k(\boldsymbol{\mu})\}_{k\in\mathcal M}}
\chi_{\bmu}^{XXZ} ({\textbf u}) \prod\limits_{k=0}^M
(\si_k^{-})^{e_k} \mid \Uparrow \rangle\,,
\label{bwf}
\end{equation}
where the sum is taken over $C^N_M$ strict partitions $\boldsymbol{\mu}$. The state $|\!\!\Uparrow \rangle $ in (\ref{bwf}) is the fully polarized state with all spins ``up'': $\mid\Uparrow\rangle \equiv \bigotimes \limits_{n=0}^M \mid \uparrow \rangle_n$. Besides, it is proposed to use bold-faced letters as short-hand notations for appropriate $N$-tuples of numbers: for instance, ${\textbf u}$ instead of $(u_1, \dots, u_N)$, etc. Therefore, the wave function $\chi_\bmu^{XXZ} ({\textbf u})$ in (\ref{bwf}) is of the form:
\begin{equation}
\chi_\bmu^{XXZ} ({\textbf u})\,=\,\sum_{S_{p_1, p_2, \dots , p_N}} \CA_S({\textbf u})\,u_{p_1}^{2 \mu_1} u_{p_2}^{2 \mu_2} \ldots u_{p_N}^{2 \mu_N}\,,
\label{xxzwf}
\end{equation}
where summation goes over all elements of the symmetric group of permutations $S_{p_1, p_2, \dots , p_N}\equiv S  \Bigl(\begin{matrix}1, & 2, & \dots , & N \\ p_1, & p_2, & \dots, & p_N \end{matrix} \Bigr)$. The amplitude $\CA_S$ is given by the product:
\begin{equation}
\CA_S ({\textbf u})\,\equiv\,\prod_{1\le j<i\le N}\frac{1-2\Dl u_{p_i}^2 + u_{p_i}^2 u_{p_j}^2}{u_{p_i}^2 - u_{p_j}^2}\,.
\label{ampxxz}
\end{equation}

The state-vectors (\ref{bwf}) are the eigen-states of the Hamiltonian (\ref {xxzham}), \[\hat H_{XXZ}\mid\!\Psi_N({\textbf u})\rangle\,=\,E_N({\textbf u})\mid\!\Psi_N({\textbf u})\rangle\,,\] if and only if the variables $u_l$ ($1\le l\le N$) satisfy the \textit{Bethe equations}
\begin{equation}
u_l^{2(M+1)}=(-1)^{N-1}\prod_{k=1}^N\frac{1-2\Dl u_l^2+u_l^2 u_k^2}{
1-2\Dl u_k^2 + u_l^2 u_k^2}\,.
\label{xxzbethe}
\end{equation}
The corresponding eigen-energies are given by
\begin{equation}
E_N({\textbf u})\,=\,- \frac12\sum_{i=1}^N(u_i^2+u_i^{-2}-2\Dl)\,.
\label{een}
\end{equation}

\subsection{Outline of the limiting models and the problem}

In our paper we shall consider two special cases of the $XXZ$ model, namely the $\Dl\to0$ and $\Dl\to -\infty$ limits. The $\Dl\to 0$ limit known as the $XX$ Heisenberg chain is a most popular and studied one \cite{lieb, niem, iz1, iz11, iz2, iz3, iz4, mal1, mal2}. The Hamiltonian of the $XX$ model describes the nearest-neighbor interactions of spin ``up'' $\mid\uparrow\rangle$ and spin ``down'' $\mid\downarrow \rangle$ states located on sites of the periodic chain in zero magnetic field as follows:
\begin{equation}
\hat H_{XX}\equiv -\,\frac 12\sum_{k=0}^M (\si_{k+1}^{-}\si_k^{+} +
\si_{k+1}^{+}\si_k^{-})\,.
\label{anis1}
\end{equation}
It is crucial that the system described by the Hamiltonian (\ref{anis1}) is equivalent to  free fermions \cite{iz1}. The case of the $XX$ magnet can also be deduced by taking the limit of infinite on-site repulsion in the boson Hubbard model (the ``hard-core'' bosons) \cite{sach}. Therefore, the $XX$ model is interesting for investigating the (quantum) phase diagram of the Hubbard model, as well as for description of the Frenkel excitons \cite{iz5}. In the last few years, the $XX$ model has attracted attention in connection with the quantum information and computation theory \cite{vk2, vk3}.

The state-vector of the $XX$ model and the correspondent Bethe equations are obtained at $\Dl =0$ from (\ref{xxzwf}) and (\ref{xxzbethe}), respectively. Up to an irrelevant pre-factor, the wave functions of the model are equal to
\begin{equation}
\chi_\bmu^{XX}({\textbf u})=\det(u_j^{2\mu_k})_{1\leq j,
k \leq N}\,\prod_{1\leq n < l \leq N}(u_l^2-u_n^2)^{-1},
 \label{xxwf}
\end{equation}
and the Bethe equations are \cite{iz1}:
\begin{equation}
u_j^{2(M+1)}=(-1)^{N-1}\,, \quad 1\le j \le N.
\label{bethe}
\end{equation}
The substitution $u^2_j=e^{i\ta_j}$ brings  these equations to the exponential form,
\begin{equation}
e^{i (M+1)\ta_j}=(-1)^{N-1},
\label{betheexp}
\end{equation}
with the solutions:
\begin{equation}
\theta_j=\frac{2\pi }{M+1}\Bigl(I_j-\frac{N-1}{2}\Bigr),
\label{besol}
\end{equation}
where $I_j$ are integers or half-integers depending on whether $N$ is odd or even. It is sufficient to consider a set of $N$ different numbers $I_j$ satisfying the condition: $M\geq I_1>I_2> \dots>I_N\geq 0$. The notation $\bth$ for $N$-tuple $(\ta_{1}, \ta_{2}, \dots, \ta_{N})$ of solutions (\ref{besol}) will be especially convenient for usage below in order to stress that one is concerned with the solution of the Bethe equation. Otherwise, it is appropriate to use ${\textbf u}$ as an indication that arbitrary set of parameters is meant. It follows from (\ref{een}) that the eigen-energies of the $XX$ model are equal to
\begin{equation}
E^{XX}_N(\bth)\,=\,- \sum_{j=1}^N\cos\ta_j\,=\,- \sum_{j=1}^N\cos\left(\frac{2\pi }{M+1} \Bigl(I_j-\frac{N-1}{2}\Bigr)\right).
\label{egen}
\end{equation}
The ground state of the model corresponds to the following solutions of the Bethe equations:
\begin{equation}
\theta_j=\frac{2\pi }{M+1}\Bigl(N-j-\frac{N-1}{2}\Bigr)\,, \quad 1\le j \le N.
\label{grstxx}
\end{equation}

Less studied limit of the $XXZ$ model is the \textit{Strong Anisotropy} (SA) limit $\Dl\to-\infty$ \cite{yy2, god, alc, abar, ess1, karb}. In this limit the behavior of the system is described by the effective Hamiltonian which is formally equivalent to the $XX$ Hamiltonian  supplied with requirement forbidding two spin ``down'' states to occupy any pair of nearest-neighboring sites \cite{alc, abar}:
\begin{equation}
\hat H_{\inf} = -\,\frac{1}{2}\,\sum_{k=0}^M \CP ( \si_{k+1}^{-}
\si_k^{+}+\si_{k+1}^{+}\si_k^{-})\CP\,,
\label{anis2}
\end{equation}
where the projector $\CP$ cuts out the states with the spin ``down'' states at any pair of nearest-neighboring sites: $\CP\equiv \displaystyle{ \prod_{k=0}^M}(1-\hat q_{k+1} \hat q_k)$. The local projectors onto the spin ``up'' and ``down'' states are equal to:
\begin{equation}
\check q_k\,\equiv\,\frac12\,(\si^0_k + \si^z_k)\,,\quad \hat q_k\,\equiv\,\frac12\,(\si^0_k - \si^z_k)\,,\quad \check q_k + \hat q_k\,=\,\BI , \qquad k\in\CM\,,
\label{proj}
\end{equation}
where the operators $\si^{\#}_k$ are defined by (\ref{xxzham1}).

In the limit $\Dl\to-\infty$, the wave function (\ref{xxzwf}) takes the form:
\begin{equation}
\chi_\bmu^{\inf}({\textbf u})=\det(u_j^{2(\mu_k-N+k)})_{1\leq j, k\leq N}\,\prod_{1\leq n<l\leq N}(u_l^2-u_n^2)^{-1}\,,
 \label{xxwf1}
\end{equation}
where the coordinates of the spin ``down'' states form strict decreasing partition $\bmu$ (as in (\ref{xxzwf}) and (\ref{xxwf})), i.e., $M\geq\mu_1>\mu_2>\,\dots\,> \mu_N\geq 0$. It follows from (\ref{xxwf1}) that the wave function is not equal to zero if and only if the elements $\mu_i$, $1\le i\le N$, satisfy the \textit{exclusion condition}: $\mu_i>\mu_{i+1}+1$. It is crucial that in the considered limit the occupation of nearest sites is forbidden, and the \textit{hard-core diameter} equal to duplicated inter-site separation arises. The Bethe equations of the model take the form:
\begin{equation}
e^{i (M+1-N) \theta_k}=(-1)^{N-1}\prod_{j=1}^N e^{-i\theta_j}\,, \quad 1 \le k\le N\,,
\label{bthv}
\end{equation}
and have the  solutions
\begin{equation}
\theta_j=\frac{2\pi}{M+1-N}\Bigl(I_j-\frac{N-1}{2}-P\Bigr),
\label{sol}
\end{equation}
where $P \equiv \displaystyle{\frac{1}{2\pi}\sum\limits_{j=1}^N \theta_j}$, and $I_j$ are integers or half-integers depending on $N$ being odd or even,  satisfying the  condition $M-N\geq I_1 > I_2 > \dots> I_N\geq 0$. The ground state of the model is defined by the solutions
\begin{equation}
\ta_j=\frac{2\pi }{M+1-N}\Bigl(N-j-\frac{N-1}{2}\Bigr),
\label{grstsa}
\end{equation}
while $P=0$ is fulfilled. The eigen-energy of the model is:
\begin{equation}
E_N(\bth)\,=\,-\sum_{j=1}^N\cos\left(\frac{2\pi}{M+1-N} \Bigl(I_j-\frac{N-1}{2}-P\Bigr)\right) \,.
\label{eigenv}
\end{equation}

Similarity between these two limits, (\ref{anis1}) and (\ref{anis2}), is that their wave functions are expressible through the {\it Schur functions} \cite{macd}:
\begin{equation}
\begin{array}{r}
S_{\bla} ({\textbf x})\,\equiv\,
\displaystyle{
S_{\bla} (x_1, x_2, \dots, x_N) \,\equiv\,\frac{\det(x_j^{\la_k+N-k})_{1\leq j,k\leq
N}}{\det(x_j^{N-k})_{1\leq j, k\leq N}}}  \\ [0.6cm]
  =\,\displaystyle{\det(x_j^{\la _k+N-k})_{1\leq j, k\leq N}\,\prod_{1\leq n<l\leq N}(x_l-x_n)^{-1}}\,,
\end{array}
\label{sch}
\end{equation}
where ${\bla}$ denotes the partition $(\la_1, \la_2, \dots, \la_N)$ being $N$-tuple of non-increasing non-negative integers:
$L\geq\la_1\ge\la_2\ge\,\dots\,\ge \la_N\geq 0$. Indeed, any strict partition $M\geq\mu_1>\mu_2>\,\dots\,> \mu_N\geq 0$ and non-strict partition $M+1-N\geq \la_1\geq \la_2\geq \dots\geq \la_N\geq 0$ (denoted as $\bmu$ and $\bla$, respectively) can be related by means of the relation $\la_j=\mu_j-N+j$, where $1\le j\le N$.
In other terms, $\bla =\bmu -\bdl$, where $\bdl$ is the strict partition $(N-1, N-2, \dots, 1, 0)$. So the wave function of the $XX$ model (\ref{xxwf}) may be represented as
\begin{equation}
\chi_{\bmu}^{XX} ({\textbf u})=S_{\bla} ({\textbf u}^2)\,.  \label{csh}
\end{equation}
Any strict partition $\bmu$ with the elements respecting the exclusion condition $\mu_i>\mu_{i+1}+1$ is connected with the  non-strict partition $\widetilde{\bla}$ by the relation $\widetilde{\bla} =\bmu -2\bdl$, where $M+2(1-N)\geq \widetilde \la_1 \geq \widetilde\la_2\geq \dots\geq \widetilde \la_N\geq 0$. Therefore, Eq. (\ref{xxwf1}) is re-expressed:
\begin{equation}
\chi_{\bmu}^{\inf} ({\textbf u})=S_{\widetilde{\bla}}
({\textbf u}^2)\,.
\label{cshinf}
\end{equation}

It is useful to remind a graphical picture (see Figure 1) of the correspondence between strict partitions $\bmu$ and non-strict partitions $\bla$ used in (\ref{csh}).
Namely, to each partition $\bla$ we associate a set of $N$-tuples $\mathfrak{G}(\bla)$ as follows \cite{oko}:
\[
\mathfrak{G}(\bla)\,=\,\Bigl\{\la_j-j+\frac12\,\Bigl|\, 1\le j\le N \Bigr\}\subset \BZ\,+\,\frac12\,.
\]
On another hand, any non-strict partition $\bla$ can be represented as a rectangular table (the Young table) consisting of $N$ columns so that  $\la_i$, $\forall i$, is the height of $i^{th}$ column ($\la_i \le M-N+1$).
\begin{figure}[h]
\centering
\includegraphics{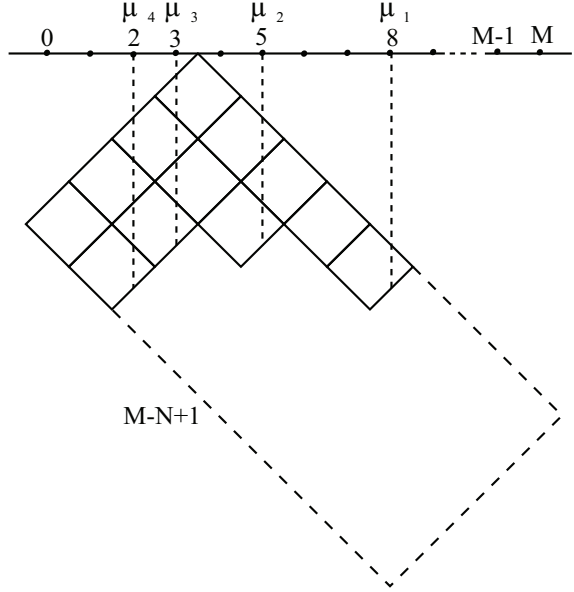}
\caption{
The strict partition $(\mu_1, \mu_2, \mu_3, \mu_4)$ and
the corresponding Young table.}
\end{figure}
We shift each element of the set $\mathfrak{G}(\bla)$ by $N+\frac12$. Then we assign the numbers thus obtained to the projections along the vertical dashed lines onto the horizontal axis. The set of points on the horizontal axis just provides the strict partition $\bmu$. For instance, the diagram on Figure~1 is drown for $M=8$ and $N=4$, and we have got respectively: $\bla=(5, 3, 2, 2)$ and $\bmu=(8, 5, 3, 2)$.

In our paper we shall study the thermal correlation function of the states (would be called as {\textit {ferromagnetic strings}}) with no spins down on the last $n+1$ sites of the lattice. We are going to consider the expectation value defined by the ratio:
\begin{equation}
\CT ({\bth}, n, \be)\,\equiv\,\frac{\langle
\Psi_N({\bth})\mid \bar\varPi_{n}\,e^{-\be \hat H}\,
\bar\varPi_{n} \mid\!\Psi_N({\bth})\rangle }{\langle
\Psi_N({\bth})\mid\!\Psi_N({\bth})\rangle}\,,\qquad \bar\varPi_{n} \equiv \prod\limits_{j=M-n}^M \check q_j\,,
\label{ratbe0}
\end{equation}
where $\be\in\BC$, and the projector $\bar\varPi_{n}$ is expressed by means of ${\check q}_j$ (\ref{proj}). Besides, $\hat H$ in $\CT ({\bth}, n, \be)$ (\ref{ratbe0}) implies either $\hat H_{XX}$ (\ref{anis1}) or $\hat H_{\inf}$ (\ref{anis2}), and $\bth$ indicates that the eigen-state $\mid\!\Psi_N({\bth})\rangle$ is calculated for solution of the Bethe equation (\ref{betheexp}) or (\ref{bthv}), respectively. Our calculations will be based on the similarity of the problem under consideration with the problem of enumeration of admissible lattice paths made by random vicious walkers. We shall extensively use the relation for the Schur functions (\ref{sch}) which is due to the Binet--Cauchy formula \cite{gant}:
\begin{eqnarray}
&& \sum_{\bla \subseteq \{L^N\}}S_{\bla} (x_1^2, \dots,
x_N^2) S_{\bla} (y_1^2, \dots, y_N^2)  \label{schr} \\
  &=& \det (T_{jk})_{1\le j, k\le N} \prod_{1\leq k<j \leq N} \left( y_j^2-y_k^2\right)^{-1} \prod_{1\leq m<l\leq N} \left(x_l^2-x_m^2\right)^{-1}\,.  \nonumber
\end{eqnarray}
The entries of the matrix $T_{jk}$ take the form:
\begin{equation}
T_{jk}=\frac{1-(x_k y_j)^{2(N+L)}}{1-(x_k y_j)^2}\,.
\label{hm}
\end{equation}
Summation in (\ref{schr}) goes over all non-strict partitions $\bla$ into at most $N$ parts so that each is less than $L$: $L\ge \la_1 \ge \la_2\ge \dots \ge \la_N \ge 0$. The notation for the range of summation in (\ref{schr}) will be extensively used in the rest of the paper.

\section{XX Heisenberg chain}

\subsection{The Bethe states and form-factors}

Before to proceed with calculation of $\CT ({\bth}, n, \be)$ (\ref{ratbe0}), let us apply our approach to more familiar examples. First of all, let us specialize, with regard at (\ref{csh}), the state-vector (\ref{bwf}) and its conjugated as follows:
\begin{equation}
\begin{array}{l}
\mid\!\Psi_N({\textbf u})\rangle = \sum\limits_{\bla \subseteq \{(M+1-N)^N\}} S_\bla ({\textbf u}^{2}) \prod\limits_{k=0}^M (\si_k^{-})^{e_k} \mid \Uparrow \rangle\,,\\ [0.5cm]
\langle \Psi_N({\textbf v})\!\mid = \sum\limits_{\bla \subseteq \{(M+1-N)^N\}} \langle \Uparrow \mid \prod\limits_{k=0}^M (\si_k^{+})^{\tilde e_k} S_\bla ({\textbf v}^{-2}) \,,
\end{array}
\label{conwf1}
\end{equation}
where summation goes over all non-strict partitions $\bla$, which are related to the non-strict par\-ti\-tions $\bmu=\bla+\bdl$. Because of the ortho\-go\-na\-li\-ty relation
\[
\langle \Uparrow \mid \prod\limits_{k=0}^M (\si_k^{+})^{\tilde
e_k} \prod\limits_{l=0}^M (\si_l^{-})^{e_l}| \!\!\Uparrow \rangle
=\prod\limits_{n=0}^M \dl_{{\tilde e}_n e_n}\,,
\]
the scalar product of the state-vectors (\ref{conwf1}) takes the form:
\begin{equation}
\langle \Psi_N({\textbf v})\mid\!\Psi_N({\textbf u})\rangle\,=\,
\sum_{\bla \subseteq \{(M+1-N)^N\}}S_\bla ({\textbf
v}^{-2})S_\bla ({\textbf u}^2)\,=\,\frac{\det(T_{k j})_{1\leq
k, j\leq N}}{{\CV}({\textbf u}^2){\CV}({\textbf v}^{-2})}
\,.
\label{spxx}
\end{equation}
We use in (\ref{spxx}) the notation for the Vandermonde determinant,
\begin{equation}
\CV({\textbf u}^{2})\,\equiv\,\prod_{1\leq m<l\leq N}(u_l^2-u_m^2)\,, \label{spxx1}
\end{equation}
while the entries $T_{k j}$ of the matrix of the size $N\times N$ are of the form:
\begin{equation}
T_{kj}=\frac{1-(u^2_k/v^2_j)^{M+1}}{1-u^2_k/v^2_j}\,.
\label{t}
\end{equation}
Equations (\ref{spxx}) and (\ref{t}) are specifications of Eqs. (\ref{schr}) and (\ref{hm}), respectively. On the solutions (\ref{besol}), the entries (\ref {t}) are equal to $T_{jk}=(M+1)\dl_{jk}$, where l'Hospital rule is taken into account on the principle diagonal. Let us use the exponential parametrization for the solutions of the Bethe equations in compact form:
\begin{equation}
{\textbf u}^{2}=e^{i \bth}\,,\qquad  e^{i \bth} \equiv (e^{i \ta_{1}}, e^{i \ta_{2}}, \dots, e^{i \ta_{N}})
\,,
\label{param}
\end{equation}
where the ``angle'' notation (\ref{besol}) or (\ref{sol}) is meant. Thus, the answer for squared norm $\CN^2({\bth})\equiv \langle \Psi_N({\bth}) \mid\!\Psi_N ({\bth}) \rangle$ of the Bethe eigen-vector (\ref{conwf1}) arises as follows:
\begin{equation}
\CN^2({\bth})\,=\,\displaystyle{\frac{(M+1)^N}{\CV(e^{i\bth}) \CV(e^{-i\bth})} =\frac{(M+1)^N}{\prod\limits_{1\leq m<l\leq N} 2 (1-\cos (\theta_{l}-\theta_{m}))}}\,.
\label{normxx}
\end{equation}

Let us now turn to the ratio:
\begin{equation}
\CT ({\textbf v}, {\textbf u}, n)\,\equiv\,\frac{\langle \Psi_N({\textbf v})\mid \bar\varPi_{n}\mid\!\Psi_N({\textbf u})\rangle }{\CN({\textbf v})\CN({\textbf u})}\,,
\label{ratio}
\end{equation}
where the projector $\bar\varPi_{n}$ is defined in (\ref{ratbe0}), and $\CN^2({\textbf u})=\langle \Psi_N({\textbf u})\mid\!\Psi_N({\textbf u})\rangle$ for arbitrary parametrization of the state-vectors. With regard at (\ref{conwf1}), we calculate:
\begin{equation}
\bar\varPi_{n}\,\mid\!\Psi_N ({\textbf u})\rangle\,=\,\sum_{{\bla}
\subseteq \{(M-N-n)^N\}} S_{{\bla}}({\textbf
u}^2)\,\Bigl(\prod\limits_{k=M-n}^M
(\si^-_k)^0\Bigr)\,\Bigl(\prod\limits_{k=0}^{M-n-1}
(\si^-_k)^{e_k}\Bigr) \mid\Uparrow\rangle \,,
\label{dfpxx1}
\end{equation}
where summation goes over non-strict partitions
${\bla}$ respecting the condition: $M-N-n\ge\la_1 \ge\la_2 \ge\,\dots\, \ge \la_N\ge 0$. Taking into account (\ref{conwf1}) and (\ref{dfpxx1}), and using the Binet-Cauchy formula, we calculate the nominator of (\ref{ratio}):
\begin{eqnarray}
\langle \Psi_N({\textbf v})\mid\bar\varPi_{n}\mid\!\Psi_N({\textbf u})\rangle\,=\,\sum_{{\bla} \subseteq \{(M-N-n)^N\}}S_{\bla} ({\textbf v}^{-2}) S_{\bla} ({\textbf u}^2) &&  \nonumber \\
=\displaystyle{ \frac{1}{{\CV}({\textbf u}^2){\CV}({\textbf
v}^{-2})} \det\Biggl(\frac{1 - (u_k^2/v_j^2)^{M-n}}{1 -
u_k^2/v_j^2}\Biggr)_{1\le j, k \le N}}\,.&&
\label{spdfpxx}
\end{eqnarray}

Assume that the sets of the parameters ${\textbf v}$ and ${\textbf u}$ in (\ref{ratio}) coincide and consist of the solutions (\ref{besol}). Then, expression $\CT ({\bth}, n) \equiv \CT (e^{i \bth/2}, e^{i \bth/2}, n)$ is related to, so-called, \textit{Emptiness Formation Probability}, which provides the probability of formation of a string of the empty (i.e., spin ``up'') states on last $n+1$ sites of the lattice. Eventually, Eq. (\ref{spdfpxx}) leads to $\CT ({\bth}, n)$ in the same determinantal form as in \cite{iz1, iz2} (see therein for more Refs.):
\begin{equation}
\CT ({\bth}, n)\,=\,
\det\Bigr(\bigl(1-\frac{n+1}{M+1}\bigr)\dl_{jk}+\frac{1-e^{i (\ta_{j}-\ta_{k})(n+1)}} {(M+1)(1-e^{i(\ta_{k}-\ta_{j})})}
(1-\dl_{j k})\Bigl)_{1\le j, k \le N}\,, \label{adfpxx}
\end{equation}
where the parameters $\ta_{l}$, $1\le l\le N$, correspond to the parametrization (\ref{besol}).

\subsection{Thermal correlator of ferromagnetic string and random walks of vicious walkers}

Let us turn to calculation of the following expectation value:
\begin{equation}
\CT ({\textbf v}, {\textbf u}, n, \be)\,\equiv\,\frac{\langle
\Psi_N({\textbf v})\mid \bar\varPi_{n}\,e^{-\be \hat H_{XX}}\,
\bar\varPi_{n} \mid\!\Psi_N({\textbf u})\rangle }{\CN({\textbf v})\CN({\textbf u})}\,,
\label{ratbe}
\end{equation}
which is clearly reduced to $\CT ({\textbf v}, {\textbf u}, n)$ (\ref{ratio}) at $\be=0$. However, our aim is to obtain $\CT ({\bth}, n, \be)$ (\ref{ratbe0}) just taking the arguments ${\textbf u}$ and ${\textbf v}$ in (\ref{ratbe}) coinciding with the same solution (\ref{besol}).

Let us use the technique presented above to calculation of the nominator of (\ref{ratbe}). Using (\ref{dfpxx1}) (and its conjugated), one gets:
\begin{equation}
\begin{array}{rcl}
&&\langle \Psi _N({\textbf v})\mid \bar\varPi_{n}\, e^{-\be \hat H_{XX}}\,\bar\varPi_{n}\mid\!\Psi_N({\textbf u})\rangle \\[0.5cm]
&&=\sum\limits_{{\bla^L},\,{\bla^R} \subseteq \{(M-N-n)^N\}}
S_{{\bla^L}}({\textbf v}^{-2}) S_{{\bla^R}}({\textbf u}^2)
F_{{\bmu^L};\,{\bmu^R}} (\be)\,.
\label{ratbe1}
\end{array}
\end{equation}
Summations in (\ref{ratbe1}) run over non-strict partitions ${\bla}^L$ and ${\bla}^R$ of the same kind as in (\ref{dfpxx1}). Superscripts $L$ and $R$ are only to distinguish two independent summations. The corresponding strict partitions ${\bmu}^L$ and ${\bmu}^R$ are defined as follows: ${\bmu}^{L, R}={\bla}^{L, R}+{\bdl}$, where ${\bdl}\equiv (\dl_1, \dl_2, \dots, \dl_N)$, $\dl_j=N-j$. The notation $F_{{\bmu^L};\,{\bmu^R}}(\be)$ implies the following average:
\begin{equation}
\begin{array}{rcl}
F_{{\bmu^L};\,{\bmu^R}}(\be)&\equiv&
F_{\mu^L_1, \mu^L_2,\dots, \mu^L_N; \mu^R_1, \mu^R_2, \dots, \mu^R_N}({\be})\,=\\[0.5cm]
&=&\langle\Uparrow |\si_{\mu^L_1}^{+}\si_{\mu^L_2}^{+}\dots
\si_{\mu^L_N}^{+}e^{-\be \hat H_{XX}}\si_{\mu^R_1}^{-}
\si_{\mu^R_2}^{-}\dots\si_{\mu^R_N}^{-}|\Uparrow\rangle\,,
\label{ratbe2}
\end{array}
\end{equation}
which is nothing but $2N$-point correlation function over the ferromagnetic state. It is related to enumeration of admissible trajectories which are traced by $N$ vicious walkers traveling over sites of one-dimensional chain \cite{b1, b2, b3, b4}.

Indeed, let $|P_K ({\mu^R_1}, \dots, {\mu^R_N}\rightarrow {\mu^L_1}, \dots, {\mu^L_N})|$ be a number of trajectories consisting of $K$ links made by $N$ vicious walkers in the random turns model. Here, the initial and final positions of the walkers on the sites are  given respectively by elements of the strict decreasing partitions ${\mu^R_1} > {\mu^R_2}> \dots > {\mu^R_N}$ and ${\mu^L_1} > {\mu^L_2} >\dots > {\mu^L_N}$. We introduce the notation~$\mathcal D^K_{\ell}$ for the operator of differentiation of $K^{th}$ order with respect to~$\ell$ at
the point $\ell=0$ \cite{b4}. Then, the application of the operator $\mathcal D^K_{\be/2}$ to the correlator \eqref{ratbe2} results in the average of the type
\[
\langle\Uparrow\!\!|\si_{\mu^L_1}^{+}\si_{\mu^L_2}^{+}\dots
\si_{\mu^L_N}^{+}(-2{\hat H}_{XX})^K \si_{\mu^R_1}^{-}
\si_{\mu^R_2}^{-}\dots\si_{\mu^R_N}^{-} |\!\!\Uparrow\rangle\,.
\]
This average provides the numbers $|P_K ({\mu^R_1}, \dots, {\mu^R_N}\rightarrow {\mu^L_1}, \dots, {\mu^L_N})|$, as it has been established in \cite{b2} with the help of the commutation relation
\begin{equation}
\lbrack {\hat H}_{XX},\si_{\mu^R_1}^{-}
\si_{\mu^R_2}^{-}\dots\si_{\mu^R_N}^{-}]=
\sum_{k=1}^N\sigma_{\mu^R_1}^{-}\dots\sigma_{\mu^R_{k-1}}^{-}
[{\hat H}_{XX},\sigma_{\mu^R_{k}}^{-}]\sigma_{\mu^R_{k+1}}^{-}
\dots\sigma_{\mu^R_{N}}^{-}\,.
\label{ratbe21}
\end{equation}
The condition of non-intersection of trajectories of the walkers is expressed by the vanishing of the correlation function~\eqref{ratbe2} for any pair of coinciding indices ${\mu^R_{k}}$ or ${\mu^L_{p}}$. Thus we conclude that the average (\ref{ratbe1}) turns out to be the generating function of the polynomials dependent on $2N$ variables, $u_1^2, u_2^2, \dots, u_N^2$ and $v_1^{-2}, v_2^{-2}, \dots, v_N^{-2}$, as follows:
\begin{equation}
\begin{array}{rcl}
&&\mathcal D^K_{\be/2}\,\Bigl[\langle \Psi _N({\textbf v})\mid \bar\varPi_{n}\, e^{-\be \hat H_{XX}}\,\bar\varPi_{n}\mid\!\Psi_N({\textbf u})\rangle\Bigr]\,=\\[0.6cm]
&&=\sum\limits_{{\bla^L},\,{\bla^R} \subseteq \{(M-N-n)^N\}}
|P_K ({\bmu^R}\rightarrow {\bmu^L})|\,S_{{\bla^L}}({\textbf v}^{-2})\,S_{{\bla^R}}({\textbf u}^2)
\end{array} \label{ratbe22}
\end{equation}
(remind that ${\bmu}^{L, R}={\bla}^{L, R}+{\bdl}$). As it is shown in \cite{b4}, the number of trajectories consisting of $K$ links, which are traced by $N$ vicious walkers on an axis, i.e, $|P_K ({\bmu^R}\rightarrow {\bmu^L})|$, is expressed through the number of trajectories of the same ``length'' $K$, which are traced by a single walker traveling over sites of $N$-dimensional lattice of infinite extension.

The correlator (\ref{ratbe2}) respects the following equation:
\begin{equation}
\begin{array}{l}
\displaystyle{\frac{d}{d\be}\,F_{\mu^L_1, \mu^L_2,\dots, \mu^L_N; \mu^R_1, \mu^R_2, \dots, \mu^R_N}({\be})}\,=\\[0.5cm]
=\,\displaystyle{\frac12\sum_{k=1}^N\bigl(F_{\mu^L_1, \mu^L_2,
\dots, \mu^L_N; \mu^R_1, \mu^R_2, \dots, \mu^R_k+1, \dots,
\mu^R_N}({\be})\,+\, F_{\mu^L_1, \mu^L_2, \dots, \mu^L_N; \mu^R_1,
\mu^R_2, \dots, \mu^R_k-1, \dots, \mu^R_N}({\be})\bigr)}\,.
\label{ratbe3}
\end{array}
\end{equation}
Equation (\ref{ratbe3}) has been considered in~\cite{b2} for the case of the periodic boundary condition with respect to the lattice argument and with the initial condition:
\[
F_{\mu^L_1, \mu^L_2,\dots, \mu^L_N; \mu^R_1, \mu^R_2, \dots,
\mu^R_N}(0)=\prod_{k=1}^N\dl_{\mu^L_k, \mu^R_k}\,.
\]
Solution to (\ref{ratbe3}) can be expressed as the determinant of the matrix $\bigr(F_{k;\,l}(\be) \bigl)_{1\leq k, l\leq N}$
\cite{b1, b2}:
\begin{equation}
F_{\mu^L_1, \mu^L_2,\dots, \mu^L_N; \mu^R_1, \mu^R_2, \dots,
\mu^R_N}({\be})\,=\, \det\bigl(F_{\mu^L_k;
\mu^R_l}({\be})\bigr)_{1\leq k, l\leq N}\,, \label{ratbe4}
\end{equation}
where the entries respect the following difference-differential equation:
\begin{equation}
\frac{d}{d\be}\,F_{k;\,l}({\be})=
\frac12\,(F_{k+1;l}({\be})\,+\,F_{k-1;l}({\be}))\,. \label{ratbe5}
\end{equation}
Similar equation can be also obtained for the fixed index $l$.

It can be checked that the transition amplitude $\langle\Uparrow|\si_{k}^{+}e^{-\be \hat H_{XX}} \si_{l}^{-}|\Uparrow\rangle$ respects (\ref{ratbe5}) \cite{b1, b2, b3, b4}. This average can be considered as the generating function of the number of walks with random turns of a single pedestrian traveling between l$^{th}$ and k$^{th}$ sites of (periodic) chain \cite{b2, b4}. Solution to (\ref{ratbe5}) can be written as the following sum:
\begin{equation}
F_{k;\,l}(\be)\,\equiv\,\displaystyle{\frac{1}{M+1}
\sum\limits_{s=0}^M e^{\be\cos\phi_s}\,e^{i \phi_s(k-l)}}\,, \label{ratbe6}
\end{equation}
where the parametrization  $\phi_s=\displaystyle{\frac{2\pi}{M+1} \Bigl(s-\frac{M}{2}\Bigr)}$ is used. The periodicity condition with respect of the lattice argument and the ``initial'' condition $F_{k;\,l}(0) = \dl_{k, l}$ are imposed.

Using (\ref{ratbe4}) and (\ref{ratbe6}), we re-express
(\ref{ratbe2}) through the Schur functions (\ref{sch}) and the Vandermonde determinants (\ref{spxx1}) as follows \cite{b1}:
\begin{equation}
\begin{array}{rcl}
F_{{\bmu^L};\,{\bmu^R}}(\be)\,=\,\displaystyle{\frac{1}{(M+1)^N}
\sum\limits_{M\ge k_1 > k_2 \dots > k_N\ge 0}
e^{\be\sum\limits_{l=1}^N\cos(\phi_{k_l})}}&& \\[0.7cm]
\times {\CV}(e^{i\bphi}){\CV}(e^{-i\bphi})\,S_{{\bla^L}}(e^{i\bphi})
S_{{\bla^R}}(e^{-i\bphi})\,.&& \label{ratbe7}
\end{array}
\end{equation}
The parametrization $\phi_s$, $0\le s\le M$, used in (\ref{ratbe7}) is the same as in (\ref{ratbe6}). We continue to use bold-faced letters to denote $N$-tuples of numbers: for instance, ${\bphi}$ corresponds to $(\phi_{k_1}, \phi_{k_2}, \dots, \phi_{k_N})$. We substitute (\ref{ratbe7}) into (\ref{ratbe1}), and use (\ref{spdfpxx}) in order to calculate the sums:
\begin{equation}
{\CP}({\textbf v}^{-2},
e^{i\bphi})\,\equiv\,\sum\limits_{{\bla^L}}S_{{\bla^L}}({\textbf
v}^{-2})\, S_{{\bla^L}}(e^{i\bphi})\,,\quad {\CP}(e^{-i\bphi},
{\textbf u}^{2})\,\equiv\, \sum\limits_{{\bla^R}}
S_{{\bla^R}}(e^{-i\bphi})\, S_{{\bla^R}}({\textbf u}^2)\,.
\label{ratbe71}
\end{equation}
The range of summation in (\ref{ratbe71}) is taken as follows: ${\bla^L},\,{\bla^R} \subseteq \{(M-N-n)^N\}$. Then we obtain:
\begin{equation}
\begin{array}{rcl}
&&\langle \Psi _N({\textbf v})\mid \bar\varPi_{n}\, e^{-\be \hat H_{XX}}\,\bar\varPi_{n}\mid\!\Psi_N({\textbf u})\rangle\,=
\\[0.6cm]
&&=\,\displaystyle{\frac{1}{(M+1)^N\,{\CV}({\textbf u}^2){\CV}({\textbf v}^{-2})}\sum\limits_{M\ge k_1 > k_2 \dots > k_N\ge 0}\!\!e^{\be\sum\limits_{l=1}^N \cos(\phi_{k_l})}\times}\\[0.6cm]
&&\times\displaystyle{ \det \Bigl(
\frac{1-(e^{i\phi_{k_i}} v_j^{-2})^{M-n}}{ 1- e^{i\phi_{k_i}} v_j^{-2}}\Bigr)_{1\le i, j \le N}\,\det
\Bigl(\frac{1-(u_p^2e^{-i\phi_{k_l}})^{M-n}}{1-u_p^2
e^{-i\phi_{k_l}}}\Bigr)_{1\le p, l \le N}}\,.
\end{array}
\label{ratbe8}
\end{equation}
The Binet--Cauchy formula enables to evaluate (\ref{ratbe8}) as follows:
\begin{equation}
\begin{array}{rcl}
&&\langle \Psi _N({\textbf v})\mid \bar\varPi_{n}\, e^{-\be \hat H_{XX}}\,\bar\varPi_{n}\mid\!\Psi_N({\textbf u})\rangle\,=\\[0.6cm]
&&=\,\displaystyle{ \frac{1}{{\CV}({\textbf u}^2){\CV}({\textbf
v}^{-2})} \det \left(\sum\limits_{k, l=0}^{M-n-1}
F_{k;\,l}(\be)\,\frac{u_i^{2l}}{v_j^{2k}}\right)_{1\le i,j \le
N}}\,,\end{array} \label{ratbe9}
\end{equation}
where $F_{k;\,l}(\be)$ is defined by (\ref{ratbe6}).
Clearly, the relation (\ref{ratbe9}) at $\be=0$ is reduced to (\ref{spdfpxx}). Expression for $\CT ({\bth}, n, \be)$ (\ref{ratbe0}) can straightforwardly be obtained from (\ref{ratbe9}) at coinciding parameters ${\textbf u}={\textbf v}$ (being solutions of the Bethe equations) with the help of $\CN^2$ (\ref{normxx}):
\begin{equation}
\CT ({\bth}, n, \be)\,=\,\displaystyle{ \frac{1}{(M+1)^{N}}\,\det
\Biggl(\sum\limits_{k, l=0}^{M-n-1}
F_{k;\,l}(\be)\,e^{i(l\ta_i-k\ta_j)}\Biggr)_{1\le i,j \le N}}\,.
\nonumber
\end{equation}

\section{Strongly anisotropic ${\textbf {XXZ}}$ chain}

\subsection{The Bethe state vectors and their form-factors}

Let us turn to the strong anisotropy limit, $\Dl\rightarrow -\infty$, which is described by the Hamiltonian $\hat H_{\inf}$ (\ref{anis2}). The corresponding state-vector is given by Eq. (\ref{bwf}),
\begin{equation}
\mid\!\Psi_N({\textbf u})\rangle = \sum_{\widetilde\bla \subseteq \{(M-2(N-1))^N\}} S_{\widetilde\bla}
({\textbf u}^2)\prod\limits_{k=0}^M (\si_k^{-})^{e_k}\mid\Uparrow\rangle\,,
\label{vstv}
\end{equation}
where $S_{\widetilde\bla} ({\textbf u}^2)$ is given by (\ref{cshinf}), and summation goes over all non-strict partitions. Summation over strict partitions $\bmu$ with the elements respecting the condition $\mu_i>\mu_{i+1}+1$ is equivalent to that over the non-strict partitions $\widetilde{\bla}$, where $\widetilde{\bla} =\bmu -2\bdl$ and  $M+2(1-N)\geq \widetilde \la_1 \geq \widetilde\la_2\geq \dots\geq \widetilde \la_N\geq 0$.
The scalar product of the state-vector (\ref{vstv}) to its conjugated (defined similarly to (\ref{conwf1})) is given by the relation looking, in turn, similarly to (\ref{spxx}):
\begin{equation}
\langle \Psi _N({\textbf v})\mid\!\Psi_N({\textbf u})\rangle \,=\,\sum_{\widetilde\bla \subseteq \{(M-2(N-1))^N\}}S_{\widetilde\bla}
({\textbf v}^{-2}) S_{\widetilde\bla} ({\textbf u}^2)\,=\, \frac{\det(T_{k j})_{1\leq k, j\leq N}}{{\CV}({\textbf u}^2){\CV} ({\textbf v}^{-2})}\,,
\label{spv}
\end{equation}
where the entries $T_{k j}$ take the form:
\begin{equation}
T_{k j}=\frac{1-(u^2_k/v^2_j)^{M-N+2}}{1-u^2_k/v^2_j}\,,
\label{tv}
\end{equation}
and the notation (\ref{spxx1}) for the Vandermonde determinant is used. When the sets of parameters ${\textbf v}$ and ${\textbf u}$ in left-hand side of (\ref{spv}) coincide, usage of the Bethe equations (\ref{bthv}) enables to express the entries (\ref{tv}) as follows: $T_{j k}=1 + (M-N+1) \dl_{j k}$. Therefore the norm of the Bethe eigen-vectors ${\CN}^2({\bth}) \equiv \langle \Psi_N({\bth})\mid\!\Psi_N({\bth})\rangle$ is given by
\begin{equation}
{\CN}^2(\bth)\,=\,\displaystyle{\frac{(M+1)(M+1-N)^{N-1}}{\CV(e^{i\bth}) \CV(e^{-i\bth})} \,=\,\frac{(M+1)(M+1-N)^{N-1}}{\prod\limits_{1\leq m<l\leq N}2(1-\cos(\ta_{l}-\ta_{m}))}} \,.
\label{normv}
\end{equation}
The exponential parametrization (\ref{sol}) is meant in (\ref{normv}) in the compact form (\ref{param}).
It can be shown that the scalar product $\langle
\Psi_N({\textbf v})\!\mid\!\Psi_N({\textbf u}) \rangle$ (\ref{spv}) vanishes (i.e., the state-vectors are or\-tho\-go\-nal) provided the parameters ${\textbf u}$ and ${\textbf v}$ are independent Bethe solutions.

The nominator of the ratio (\ref{ratio}) is calculated in the same way as in the Section 2.1, i.e., by means of the Binet-Cauchy formula:
\begin{eqnarray}
\langle \Psi_N({\textbf v})\mid\bar\varPi_{n}\mid\!\Psi_N({\textbf u})\rangle\,=\,\sum_{{\widetilde\bla} \subseteq \{(M-2 N-n+1)^N\}}S_{\widetilde\bla} ({\textbf v}^{-2}) S_{\widetilde\bla} ({\textbf u}^2) &&  \nonumber \\
=\displaystyle{
\frac{1}{{\CV}({\textbf u}^2){\CV}({\textbf v}^{-2})} \det\Bigl(\frac{1 - (u_k^2/v_j^2)^{M-N-n+1}}{1 - u_k^2/v_j^2}\Bigr)_{1\le j, k \le N}}\,.&&
\label{spdfpxx2}
\end{eqnarray}
After use of (\ref{normv}) and (\ref{spdfpxx2}), the answer for (\ref{ratio}), taken on the solutions of the Bethe equation (\ref{sol}) (i.e., the Emptiness Formation Probability), appears in the following form:
\begin{eqnarray}
\CT ({\bth}, n) = \frac{M-N+1}{M+1}\,\times\,\det \Bigl(\bigl(1-\frac{n}{M-N+1}\bigr)\dl_{jk}\,+&& \nonumber\\
+\,\frac{1-e^{i n (\ta_{j}-\ta_{k})}
}{(M-N+1)(1-e^{i(\ta_{k}-\ta_{j})})} (1-\delta_{j k})\Bigr)_{1\leq k, j\leq N}\,.&&
\label{dfpv}
\end{eqnarray}

\subsection{Thermal correlator of ferromagnetic string}

Let us turn to obtaining of the average $\CT ({\textbf v}, {\textbf u}, n, \be)$ (\ref{ratbe}), where the Hamiltonian $\hat H_{XX}$ is replaced by $\hat H_{\inf}$ (\ref{anis2}). Using (\ref{vstv}) we obtain the corresponding nominator of (\ref{ratbe}):
\begin{equation}
\begin{array}{rcl}
&&\langle \Psi _N({\textbf v})\mid \bar\varPi_{n}\, e^{-\be {\hat H}_{\inf}}\,\bar\varPi_{n}\mid\!\Psi_N({\textbf u})\rangle
\,=\\[0.6cm]
&&=\,\sum\limits_{{{\widetilde\bla}^L},\,{{\widetilde\bla}^R} \subseteq \{(M-2N-n+1)^N\}} S_{{{\widetilde\bla}^L}}({\textbf v}^{-2})\,S_{{{\widetilde\bla}^R}}({\textbf u}^2)\, F_{{{\widetilde\bmu}^L};\,{{\widetilde\bmu}^R}} (\be)\,,
\end{array}\label{ratbe17}
\end{equation}
where $F_{{{\widetilde\bmu}^L};\,{{\widetilde\bmu}^R}} (\be)$ is $2N$-point correlation function over the ferromagnetic state given, practically, by (\ref{ratbe2}) excepting that $\hat H_{XX}$ (\ref{anis1}) is replaced by $\hat H_{\inf}$ (\ref{anis2}). Summations in (\ref{ratbe17}) go over non-strict partitions ${\widetilde\bla}^L$ and ${\widetilde\bla}^R$ of the same kind like in (\ref{spv}). The corresponding strict partitions ${\widetilde\bmu}^L$ and ${\widetilde\bmu}^R$ are defined as follows: ${\widetilde\bmu}^{L, R} = {\widetilde\bla}^{L, R} + 2 {\bdl}$, where ${\bdl}$ implies the partition $(N-1, N-2, \dots, 0)$. It is crucial that now the lattice indices ${\widetilde\bmu^L}$, ${\widetilde\bmu^R}$ respect the exclusion requirement: occupation of nearest sites is forbidden. Besides, an analogue of the relation (\ref{ratbe22}) can be written as follows:
\begin{equation}
\begin{array}{rcl}
&&\mathcal D^K_{\be/2}\,\Bigl[\langle \Psi _N({\textbf v})\mid \bar\varPi_{n}\, e^{-\be {\hat H}_{\inf}}\,\bar\varPi_{n}\mid\!\Psi_N({\textbf u})\rangle\Bigr]\,=\\[0.5cm]
&&=\sum\limits_{{{\widetilde\bla}^L},\,{{\widetilde\bla}^R}
\subseteq \{(M-2N-n+1)^N\}} |P_K ({{\widetilde\bmu}^R}\rightarrow
{{\widetilde\bmu}^L})|\,S_{{{\widetilde\bla}^L}}({\textbf
v}^{-2})\,S_{{{\widetilde\bla}^R}}({\textbf u}^2)\,.
\end{array} \label{ratbe214}
\end{equation}

The solutions of the Bethe equation (\ref{bthv}) constitute a complete set of the eigen-states \cite{abar}. Taking into account the orthogonality of the corresponding states, one can consider the resolution of the identity operator:
\begin{equation}
{\BI}\,=\,\sum\limits_{\{{\bth}\}} \CN^{-2}({\bth}) \mid\!\Psi_N({\bth})\rangle \langle \Psi_N({\bth})\!\mid\,,
\label{ratbe171}
\end{equation}
where summation goes over all independent solutions of the Bethe equation (\ref{bthv}), and the square of the norm $\CN^{2}({\bth})$ is given by (\ref{normv}). We shall calculate (\ref{ratbe17}) inserting (\ref{ratbe171}) into left-hand side of (\ref{ratbe17}) and using appropriately Eq.~(\ref{spdfpxx2}). We take into account that
\[\langle \Psi_N({\bf v})\mid e^{-\be \hat H_{\inf}}\mid\!\Psi_N({\bth})\rangle\,=\,\langle \Psi_N({\bf v})\mid\!\Psi_N({\bth})\rangle\,e^{-\be E_N ({\bth}) },
\]
where the expression for the energy $E_N ({\bth})$ is given by the relation (\ref{eigenv}). 
Further, we use (\ref{spdfpxx2}) and obtain:
\begin{equation}
\begin{array}{rcl}
&&\langle \Psi _N({\textbf v})\mid \bar\varPi_{n}\, e^{-\be {\hat H}_{\inf}}\,
\bar\varPi_{n} \mid\!\Psi_N({\textbf u})
\rangle \\
&=&
\displaystyle{\frac{1}{(M+1)(M+1-N)^{N-1}\,{\CV}({\textbf u}^2){\CV}({\textbf v}^{-2})} \sum\limits_{M-N\ge l_1 > l_2 \dots > l_N\ge 0}
\!\!e^{-\be E_N ({\bth})}}\\[0.6cm]
&\times&\displaystyle{\det \Bigl(
\frac{1-(e^{i\theta_{i}}v_j^{-2})^{M-N-n+1}}{1- e^{i\theta_{i}}v_j^{-2}}
\Bigr)_{1\le i, j \le N}\,\det
\Bigl(\frac{1-(u_l^2 e^{-i\theta_{p}})^{M-N-n+1}}{1-u_l^2
e^{-i\theta_{p}}}\Bigr)_{1\le l, p \le N}}\,,
\end{array}
\label{ratbe20}
\end{equation}
where summation goes over the ordered sets $\{I_k\}_{1\le k \le N}$, which parametrize the solution (\ref{sol}). Expression for $\CT ({\bth}, n, \be)$ (\ref{ratbe0}) can be obtained by means of (\ref{normv}) and (\ref{ratbe20}), where it is necessary to put ${\bf u}^2={\bf v}^2=e^{i\bth}$, while $\bth$ is the solution (\ref{grstsa}) of the Bethe equation for the ground state.

\section{Boxed plane partitions}

We shall show that the scalar products of the state vectors and the emptiness formation probability are related to the generating functions of the boxed
plane partitions.

An array $(\pi _{i,j})_{i, j\,\ge 1 }$ of non-negative integers that are non-increasing as functions of both $i$ and $j$ $(i,j\in\{1,2, \dots\}$ is called a plane partition $\bpi$ \cite{macd}. The integers $\pi_{i,j}$ are called the parts of the plane partition, and $|\bpi |=\sum\limits_{i, j\,\ge 1} \pi _{i,j}$ is its volume. Each plane partition has a three dimensional diagram which can be interpreted as a stacks of unit cubes (three-dimensional Young diagram). The height of a stack with coordinates $(i,j)$ is equal to the part of plane partition $\pi_{i,j}$. If we have $i\leq r,\,j\leq s$ and $\pi_{i,j} \leq t$ for all cubes of the plane partition, it is said that the plane partition is contained in a box with side lengths $r,s,t$. If $\pi_{i,j}>\pi_{i+1,j}$, i.e. if the parts of plane partition $\bpi$ are decaying along each column, then $\bpi $ is called column strict plane partition. We shall call the partition $\bpi$ that are decaying along each column and each raw ($\pi_{i,j} > \pi_{i+1,j}$ and $\pi_{i,j}>\pi_{i,j+1}$) as the strict plane partition. The element $\pi_{1, 1}$ of the strict plane partition $\bpi$ satisfies the condition $\pi_{1, 1}\geq 2r-2$, if all $i,j\leq r$.

An arbitrary plane partition in a box $r\times r\times t$ may be transferred into a column strict plane partition in a box $r\times r\times (t+r-1)$ by adding to an array $(\pi _{i, j})_{i, j\,\ge 1 }$ the $r\times r$ matrix
\[
\bpi _{{\rm cspp}}=\left(
\begin{array}{cccc}
r-1 & r-1 & \cdots & r-1 \\
r-2 & r-2 & \cdots & r-2 \\
\vdots & \vdots &  & \vdots \\
0 & 0 & \cdots & 0
\end{array}
\right) ,
\]
which corresponds to a minimal column strict plane partition. The volumes of the column strict plane partition and correspondent plane partition are related
\begin{equation}
|\bpi_{{\rm cspp}}|=|\bpi |+\frac{N^2(N-1)}2.  \label{volcspp}
\end{equation}
An arbitrary plane partition in a box $r\times r\times t$ may be transferred into a strict plane partition in a box $r\times r\times (t+2r-2)$ by adding to an array $(\pi _{i, j})_{i, j\,\ge 1}$ the $r\times r$ matrix
\[
\bpi _{{\rm spp}}=\left(
\begin{array}{cccc}
2r-2 & 2r-3 & \cdots & r-1 \\
2r-3 & 2r-4 & \cdots & r-2 \\
\vdots & \vdots &  & \vdots \\
r-1 & r-2 & \cdots & 0
\end{array}
\right) ,
\]
which corresponds to a minimal strict plane partition. The volumes of the strict plane partition and correspondent plane partition are related
\begin{equation} |\bpi_{{\rm spp}}|=|\bpi |+N^2(N-1).
\label{volspp}
\end{equation}

The partition function of the three dimensional Young diagrams, or saying differently the generating function of plane partitions is equal to
\begin{equation}
Z(q)=\sum\limits_{\{\bpi\}} q^{|\bpi |},  \label{pf}
\end{equation}
where $q$ is a weight, and summation is performed over all plane partitions in a box. Formulas (\ref{volcspp}) and (\ref{volspp}) provide the connection between partition functions of the plane partitions of different
types. The generating functions of the column strict and strict plane partitions placed into a box $N\times N\times M$ are equal respectively to
\begin{eqnarray}
Z_{{\rm cspp}}(q) &=&q^{\frac {N^2}{2}(N-1)}\prod_{1\leq j,k\leq N}\frac{1-q^{M+1+j-k}}{1-q^{j+k-1}},  \label{gfcs} \\
Z_{{\rm spp}}(q) &=&q^{N^2(N-1)}\prod_{1\leq j,k \leq N}\frac{1-q^{M+3-j-k}}{1-q^{j+k-1}}.  \label{gfs}
\end{eqnarray}

The scalar product (\ref{spxx}) is related to the partition function of the column strict three-dimensional Young diagrams placed into $N\times N\times M$ box. Really, the parametrizations
$v_j=q^{-\frac j2}$ and $u_j=q^{\frac{j-1}2}$ give:
\begin{equation}
\begin{array}{l}
\langle \Psi_N(q^{-\frac 12}, \dots, q^{-\frac N2})\mid\!\Psi_N(1, \dots, q^{\frac{N-1}2})\rangle\,=\\[0.5cm]
=\,\sum\limits_{\bla \subseteq \{(M+1-N)^N\}}S_\bla (q, \dots, q^N)\,S_\bla (1, \dots, q^{N-1})\,= \\[0.5cm]
=\,\displaystyle{\frac{1}{{\CV}(q, \dots, q^N) {\CV}(1, \dots, q^{N-1})}
\det\Bigl(\frac{1-s^{j+k-1}}{1-q^{j+k-1}}\Bigr)_{1\le j,k \le N}}\,,
\end{array}
\label{gfs1}
\end{equation}
where $s=q^{M+1}$, and
\begin{equation}
\displaystyle{ {\CV}^{-1} (q, \dots, q^N)\,{\CV}^{-1}(1, \dots, q^{N-1})\,=\,q^{-\frac{N}6(N-1)(2N-1)}\,\prod\limits_{1\le k<j
\le N}\left(1-q^{j-k}\right)^{-2}}\,.
\label{gfs2}
\end{equation}
The determinant in (\ref{gfs1}) was calculated
in the paper \cite{kup} in connection with the alternating sign matrices enumeration problem:
\begin{equation}
\begin{array}{l}
\displaystyle{\det\Bigl( \frac{1-s^{j+k-1}}{1-q^{j+k-1}}\Bigr)_{1\le j,k \le N}} = \\[0.5cm] =\,\displaystyle{q^{\frac{N}6(N-1)(2N-1)}\prod_{1\le k<j \le N}\left( 1-q^{j-k}\right)^2 \prod_{k, j=1}^N
\frac{1-sq^{j-k}}{1-q^{j+k-1}}}\,.  \label{deter}
\end{array}
\end{equation}
Taking into account (\ref{gfcs}), (\ref{gfs2}), and (\ref{deter}), we obtain for (\ref{gfs1}):
\begin{equation}
\langle \Psi _N(q^{-\frac 12}, \dots, q^{-\frac{N}2}) \mid\!\Psi_N(1, \dots, q^{\frac{N-1}2})\rangle = q^{-\frac{N^2}2(N-1)} Z_{{\rm cspp}}(q)\,.
\label{spxxgf}
\end{equation}
Thus, Eq. (\ref{spxxgf}) reads that the scalar product of two state-vectors coincides at $q=1$ with the number of column strict partitions in a box $N\times N\times M$, i.e., with $Z_{{\rm cspp}}(1)$.

The same parametrizations $v_j=q^{-\frac j2}$ and $u_j=q^{\frac{j-1}2}$ enable to express the scalar product (\ref{spv}) corresponding to the strong anisotropy limit. The same representation is valid though $s=q^{M-N+2}$ now, while the range of summation over $\widetilde\bla$ takes the form: \[{\widetilde\bla \subseteq \{(M-2(N-1))^N\}}\,.\]
Taking into account Eq. (\ref{deter}), we now obtain:
\[
\langle \Psi_N(q^{-\frac 12}, \dots, q^{-\frac N2})\mid\!\Psi_N (1, \dots, q^{\frac{N-1}2})\rangle = q^{-N^2(N-1)} Z_{{\rm spp}}(q)\,,
\]
where $Z_{{\rm spp}}(q)$ is the generating function of strict plane partitions (\ref{gfs}). The corresponding value of the scalar product at $q=1$ coincides with the number of strict plane partitions $Z_{{\rm spp}}(1)$.

Now lets turn to the expectation value of the ferromagnetic string (\ref{spdfpxx}). In the present parametrization, we obtain (with regard at (\ref{gfs2}) and (\ref{deter})):
\begin{equation}
\begin{array}{l}
\langle \Psi_N (q^{-\frac 12}, \dots, q^{-\frac N2})\mid\bar\varPi_{n}\mid\!\Psi_N(1, \dots, q^{\frac{N-1}2})\rangle\,=\\[0.5cm]
=\,\sum\limits_{{\bla} \subseteq \{(M-N-n)^N\}}S_\bla (q, \dots, q^N)\,S_\bla (1, \dots, q^{N-1})\,=\,\displaystyle{\prod_{k, j=1}^N \frac{1-sq^{j-k}}{1-q^{j+k-1}}}\,,
\end{array}
\label{spdfpxx1}
\end{equation}
where $s=q^{M-n}$, i.e, (\ref{spdfpxx1}) differs from
(\ref{gfs1}) in the sense that $n+1$ is subtracted from $M+1$. The box containing the plane partitions is now of smaller maximal height:  $N\times N\times (M-n-1)$. The expectation value in left-hand side of (\ref{spdfpxx1})), being considered at $q\to 1$, ``counts'' the number of plane partitions of smaller height.
The same is true for the case of strong anisotropy, and (\ref{spdfpxx2}) takes the form:
\begin{equation}
\begin{array}{l}
\langle \Psi_N (q^{-\frac 12}, \dots, q^{-\frac N2})\mid\bar\varPi_{n}\mid\!\Psi_N(1, \dots, q^{\frac{N-1}2})\rangle\,=\\[0.5cm]
=\,\sum\limits_{{\widetilde\bla} \subseteq \{(M-2 N-n+1)^N\}} S_{\widetilde\bla} (q, \dots, q^N)\,S_{\widetilde\bla} (1, \dots, q^{N-1})\,=\,\displaystyle{\prod_{k, j=1}^N \frac{1-sq^{j-k}}{1-q^{j+k-1}}}\,,
\end{array}
\label{spdfpxx3}
\end{equation}
where $s=q^{M-N-n+1}$.

\section{Low temperature asymptotics}

Let us go over to the case of long enough $XX$-chain, i.e., $M \gg 1$ while $N$ is moderate: $M\gg N \gg 1$. Now the correlator $F_{k; l}(\be)$ is approximately given by the modified Bessel function instead of (\ref{ratbe6}):
\begin{equation}
F_{k; l}({\be})=I_{k-l}({\be})=\frac1{2\pi
}\int\limits_{-\pi}^{\pi}e^{{\be}\cos\phi}e^{i(k-l)\phi}d\phi\,.
\label{ratbe11}
\end{equation}
In the limit of small ``temperature'' ($1/{\be}\rightarrow 0$) and for moderate values of $m\equiv|k-l|$, we use the known asymptotics of the Bessel function and obtain:
\begin{equation}
F_{k; l}({\be})\simeq\frac{e^{{\be}}}{{\sqrt{2\pi{\be}}}}
\biggl(1-\frac{4m^2-1}{8{\be}}+\dots\biggr)\,, \label{ratbe13}
\end{equation}
i.e., the power decay is governed by the critical exponent $\xi=-1/2$.

Since the summations can be replaced by the integrations at large enough $M$, we obtain from (\ref{ratbe1}) and (\ref{ratbe7}):
\begin{equation}
\begin{array}{rcl}
\langle \Psi _N({\textbf v})\mid \bar\varPi_{n}\, e^{-\be \hat H_{XX}}\,\bar\varPi_{n}\mid\!\Psi_N({\textbf u})\rangle\,=\, \displaystyle{\frac{e^{\be N}}{N!}\prod\limits_{i=1}^N\Bigl(
\int\limits_{-\pi}^{\pi}\!\frac{d\phi_i}{2\pi}\Bigr) \,e^{-\be\sum\limits_{l=1}^N(1-\cos\phi_l)}}&& \\[0.6cm]
\times\,\displaystyle{{\CP}({\textbf v}^{-2}, e^{i\bphi}){\CP}(
e^{-i\bphi}, {\textbf u}^{2})}\,\prod_{1\leq k<l\leq N} \mid
\!e^{i\phi_k}-e^{i\phi_l}\!\mid^2\,,&&
\end{array}
\label{ratbe14}
\end{equation}
where the con\-ti\-nu\-ous integration variables $\phi_i\in [0, 2\pi]$ are due to the
change of the dis\-cre\-te va\-ri\-ab\-les $k_i\in\CM$ as follows: $\phi_{k_i} \mapsto \phi_i$, $\forall i$. Further, the integral (\ref{ratbe14}) can be approximated at $\be$ tending to infinity as follows:
\begin{equation}
\begin{array}{rcl}
\langle \Psi _N({\textbf v})\mid \bar\varPi_{n}\, e^{-\be \hat H_{XX}}\,\bar\varPi_{n}\mid\!\Psi_N({\textbf u})\rangle \simeq
\displaystyle{{\CP}({\textbf v}^{-2}, {\bf 1}){\CP}( {\bf 1}, {\textbf u}^{2})\,\frac{e^{\be N}}{(2\pi)^N N!} }&& \\[0.5cm]
\times\!\!\displaystyle{\int\limits_{-\infty}^{\infty}\!
\int\limits_{-\infty}^{\infty}
\cdots\int\limits_{-\infty}^{\infty}\,e^{-(\be/2)\sum\limits_{l=1}^N \phi^2_l}}\!\!\prod_{1\leq
k<l\leq N} \mid\!\phi_k-\phi_l\!\mid^2  d\phi_1 d\phi_2 \dots d\phi_N\,.&&
\end{array}
\label{ratbe15}
\end{equation}
The bold-faced argument ${\bf 1}$ in the sum, say, ${\CP}( {\bf 1}, {\textbf u}^{2})$ implies that $S_{\bla}(e^{\pm i \bphi})$ in (\ref{ratbe71}) is replaced by $S_{\bla}(1, 1, \dots, 1)$ provided the $N$-tuple of the exponentials $e^{\pm i \bphi}$ is substituted by $N$-tuple of unities. The corresponding value of the Schur function can be obtained \cite{macd}:
\begin{equation}
{\displaystyle
S_{\bla}(1, 1, \dots, 1)\,=\,\frac{\prod\limits_{1\leq k<l\leq
N}(\la_l-l-\la_k+k)}{(N-1)!(N-2)!\cdots 1!0!}}\,.
\label{rep}
\end{equation}
Right-hand side of (\ref{rep}) coincides with the dimensionality ${\sf d}_{{\bla}}\equiv \textrm{dim}\,{{\bpi}}_{\bla}$ of the unitary irreducible representations of the unitary group $\CU(N)$, which corresponds to the signature ${\bla}$, i.e., ${\sf d}_{{\bla}}=S_{\bla}(1, 1, \dots, 1)$ \cite{zhel}. The Schur functions provide a base of the ring of symmetric polynomials of $N$ variables \cite{macd}.

The integral in (\ref{ratbe15}) is the \textit{Mehta integral} \cite{meh} of the \textit{Gaussian Unitary Ensemble} of random matrices. Its value is known, and the estimate for right-hand side of (\ref{ratbe15}) acquires the form:
\begin{equation}
\langle \Psi _N({\textbf v})\mid \bar\varPi_{n}\, e^{-\be \hat H_{XX}}\,\bar\varPi_{n}\mid\!\Psi_N({\textbf u})\rangle\,\simeq\, \displaystyle{{\CP}({\textbf v}^{-2}, {\bf
1})\,{\CP}({\bf 1}, {\textbf u}^{2})}\,\frac{e^{\be N}}{\be^{N^2/2}} \prod\limits_{n=1}^{N} \frac{ \Gamma(n)}{(2\pi)^{1/2}}
\,.
\label{ratbe16}
\end{equation}
Let us consider the case when ${\textbf u}^2$ and ${\textbf v}^2$ in (\ref{ratbe16}) are the solutions of the Bethe equations (\ref{bethe}). Approximate estimates ${\textbf u}^2=e^{i\bth}\simeq{\bf 1}$ and ${\textbf v}^{-2} = e^{-i\bth}\simeq{\bf 1}$ are valid at $1\ll N\ll M$. Thus, the estimate (\ref{ratbe16}) takes the limiting form:
\begin{equation}
\begin{array}{l}
\langle \Psi_N (1, \dots, 1)\mid \bar\varPi_{n}\, e^{-\be \hat H_{XX}}\,\bar\varPi_{n}\mid\!\Psi_N(1, \dots, 1)\rangle\,\simeq\\[0.3cm]
\simeq\,\Bigl(\sum\limits_{{\bla} \subseteq \{(M-N-n)^N\}} {\sf
d}_{\bla}\,{\sf d}_{\bla}\Bigr)^2\,\displaystyle{\frac{e^{\be
N}}{\be^{N^2/2}} \prod\limits_{n=1}^{N} \frac{
\Gamma(n)}{(2\pi)^{1/2}}}\,.
\end{array}
\label{spdfpxx4}
\end{equation}
The present estimate is proportional to the square of
(\ref{spdfpxx1}) at $q=1$, i.e., to the square of the number of boxed plane partitions of the size $N\times N\times (M-n-1)$. As a result, the ratio of the average $\CT ({\bth}, n, \be)$ (\ref{ratbe0}) to the partiton function of the $XX$ model $\CZ$ \cite{lieb} can be estimated as follows:
\begin{equation}
\frac{\CT ({\bth}, n, \be)}{\CZ}\,\simeq\,
{\rm{const}}\,\times\,\Bigl(\displaystyle{\prod_{k, j=1}^N
\frac{{M-n+j-k}}{{j+k-1}}}\Bigr)^2\,
\displaystyle{\frac{e^{-\be{\cal
E}^{XX}_{{\rm{0}}}}}{{\be}^{N^2/2}}}\,, \nonumber
\end{equation}
where ${\cal E}^{XX}_{{\rm{0}}}$ is the ground state energy of the $XX$ model. In the same limit, the generating function~(\ref{ratbe22}) is specified as follows:
\begin{equation}
\begin{array}{rcl}
&&\mathcal D^K_{\be/2}\,\Bigl[\langle \Psi_N (1, \dots, 1)\mid \bar\varPi_{n}\, e^{-\be \hat H_{XX}}\,\bar\varPi_{n}\mid\!\Psi_N(1, \dots, 1)\rangle\Bigr]\,=\\[0.6cm]
&&=\sum\limits_{{\bla^L},\,{\bla^R} \subseteq \{(M-N-n)^N\}} |P_K
({\bmu^R}\rightarrow {\bmu^L})|\,{\sf d}_{{\bla^L}}\,{\sf
d}_{{\bla^R}}\,,
\end{array} \nonumber
\end{equation}
where a relationship between strict and non-strict partitions is valid: ${\bmu}^{L, R}={\bla}^{L, R}+{\bdl}$ (see, for instance, (\ref{ratbe1})).

In  the case of strong anisotropy, we use (\ref{spdfpxx3}) and obtain an analogous estimate which demonstrates
the proportionality to the square of the number of strict plane partitions in a box of the size $N\times
N\times (M-n-N)$:
\begin{equation}
\begin{array}{rcl}
\displaystyle{\frac{\CT ({\bth}, n, \be)}{\CZ}} &\simeq& \displaystyle{{\rm{const}}\,\times\,\Bigl(\sum\limits_{\widetilde{\bla} \subseteq \{(M-2N-n+1)^N\}} {\sf d}_{\widetilde{\bla}}\,{\sf d}_{\widetilde{\bla}} \Bigr)^2\,\displaystyle{\frac{e^{-\be{\cal E}^{\inf}_{{\rm{0}}}}}{\be^{N^2/2}}}}\,= \\[0.7cm]
&=&{\rm{const}}\,\times\,\Bigl(\displaystyle{\prod_{k, j=1}^N
\frac{{M-N+1-n+j-k}}{{j+k-1}}}\Bigr)^2\,\displaystyle{\frac{e^{-\be{\cal
E}^{\inf}_{{\rm{0}}}}}{\be^{N^2/2}}} \,, \nonumber
\end{array}
\end{equation}
where ${\CZ}$ is the partition function of the model in the limit of strong anisotropy \cite{yy3, god}, and ${\cal E}^{\inf}_{{\rm{0}}}$ is the corresponding ground state energy. The relation (\ref{ratbe214}) is specified as follows:
\begin{equation}
\begin{array}{rcl}
&&\mathcal D^K_{\be/2}\,\Bigl[\langle \Psi _N(1, \dots, 1)\mid \bar\varPi_{n}\, e^{-\be {\hat H}_{\inf}}\,\bar\varPi_{n}\mid\!\Psi_N(1, \dots, 1)\rangle\Bigr]\,=\\[0.6cm]
&&=\sum\limits_{{{\widetilde\bla}^L},\,{{\widetilde\bla}^R}
\subseteq \{(M-2N-n+1)^N\}} |P_K ({{\widetilde\bmu}^R}\rightarrow
{{\widetilde\bmu}^L})|\,{\sf d}_{{{\widetilde\bla}^L}}\,{\sf
d}_{{{\widetilde\bla}^R}}\,,
\end{array} \nonumber
\end{equation}
where ${\widetilde\bmu}^{L, R}={\widetilde\bla}^{L, R}+2 {\bdl}$ (see (\ref{ratbe17})).

\section{Discussion}

The $XXZ$ Heisenberg chain has been considered for two specific limits of the anisotropy parameter: $\Dl\to 0$ and $\Dl\to-\infty$. The corresponding state-vectors have been expressed by means of the symmetric Schur functions. Certain expectation values and thermal correlation functions of the ferromagnetic string operators have been calculated over the base of $N$-particle Bethe states. The expectation values obtained are of the type of the emptiness formation probability. The thermal correlator of the ferromagnetic string operator is expressed through the generating function of the lattice paths of random walks of vicious walkers. The thermal correlator in question turns out to be a generating function of certain polynomials build up from the Schur polynomials. A relationship between the expectation values obtained and the generating functions of boxed plane partitions is discussed. Asymptotic estimate of the thermal expectation value of the ferromagnetic string is obtained in the limit of zero temperature for $\Dl=0$. These estimates are expressed in terms of the dimensionality of the irreducible representations of the group $\CU(N)$.

\end{document}